\documentclass[showpacs,twocolumn,aps]{revtex4-1}
\usepackage{bm}
\usepackage{epsfig}
\usepackage{amsmath}

\def\be{\begin{equation}}
\def\lan{\left\langle}
\def\ran{\right\rangle}
\def\ee{\end{equation}}
\def\barr{\begin{array}}
\def\earr{\end{array}}

\def\nn8{\\}

\def\ed{\end{document}}

\begin{document}

\title{Thermalization in one- plus two-body ensembles for dense interacting
boson systems}

 \author{N. D. Chavda$^a$, V. K. B. Kota$^b$ and V. Potbhare$^a$}
 \affiliation{$^a$Applied Physics Department, Faculty of Technology and
Engineering,\\ M.S. University of Baroda, Vadodara 390 001, India\\
$^b$Physical Research Laboratory, Ahmedabad 380 009, India}

\begin{abstract}

Employing one plus two-body random matrix ensembles for bosons,
temperature and entropy are calculated, using different definitions, as a
function of the two-body interaction strength $\lambda$ for a system with
10 bosons $(m=10)$ in five single particle levels $(N=5)$. It is found that in a
region $\lambda \sim \lambda_t$, different definitions give essentially same
values for temperature and entropy, thus defining a thermalization region. Also,
$(m,N)$ dependence of $\lambda_t$ has been derived. It is seen that $\lambda_t$
is much larger than the $\lambda$ values where level fluctuations change from
Poisson to GOE and strength functions change from Breit-Wigner to Gaussian.

\end{abstract}

\pacs{05.30.d;05.30.Jp;05.45.Mt;05.70.-a}
\maketitle
\date{}



\section{Introduction}
\label{sec:sc1}
In recent years, the study of thermalization in isolated finite many-body
quantum systems, due to inter-particle interactions, has received considerable
interest \cite{Rigol-08,Rigol-09,Rigol-11,San-10,San-12a,San-12b,Kota-11}. This
interest has arisen mainly due to major developments in experimental study of
many-particle quantum systems such as ultracold gases trapped in optical
lattices \cite{Expt}. In particular, Rigol and Santos group employed interacting
spin-$1/2$ systems (fermions and hard core bosons) on a lattice and examined
various issues such as the role of localization and chaos, statistical
relaxation, eigenstate thermalization, ergodicity principle and so on
\cite{Rigol-08,Rigol-09,Rigol-11,San-10,San-12a,San-12b}. Horoi et al. \cite{Ho-95}
and Kota and Sahu \cite{Kota-02} examined occupancies and different definitions of
entropy for $^{28}$Si and $^{24}$Mg respectively, using nuclear shell model with
realistic interactions. It was shown that the nuclei would be in the thermodynamic
regime in general, when these are away from the ground state.
Similarly, Casati's group examined in the past, Fermi-Dirac (FD) representability of
occupation numbers for a spin-$1/2$ system on a two-dimensional  lattice
\cite{Cas-01} and more recently studied thermalization in spin-$1/2$ systems,
locally coupled to an external bath, using an approach based on the
time-dependent density-matrix renormalization group method \cite{Cas-10}. Let us
add that, as emphasized by Santos et al. \cite{San-12b}, it is possible to realize
interacting spin-$1/2$ models  experimentally in optical lattices.

On the other hand, thermalization in fermionic systems has been studied in some
detail using embedded Gaussian orthogonal ensemble of one- plus random two-body
matrix ensembles [called EGOE(1+2)] \cite{Ko-01,Go-11},  i.e. random matrix
ensembles in many-fermion spaces generated by random two-body interactions in
presence of a mean-field. The EGOEs form generic models for finite isolated interacting
many-fermion systems (for boson systems these are called BEGOE with `B' for
bosons) and they model what one may call quantum many-body chaos \cite{Ko-01,Go-11}.
The role of interactions in thermalization can be investigated by varying the
interaction strength in these models. For example, for spin-less
fermion systems, Flambaum et al. showed that EGOE(1+2) exhibits a region of
thermalization \cite{Flam-97} and found the criterion for the occupancies to
follow FD distribution\cite{Flam-97,Flam-96}. Later, using EGOE
both for spin-less fermions and fermions with spin \cite{Kota-02,An-04,Ma-10},
thermalization region generated by random interactions has been established by analyzing
different definitions for entropy. Going beyond these, in a more detail study,
thermalization has been investigated within EGOE(1+2) for spin-less fermions by
Kota et al. \cite{Kota-11} using the ergodicity principle for the expectation
values of different types of operators. Recently, Santos et al.
compared results of spin-less EGOE(1+2) for statistical relaxation \cite{Fl-01}
with those from spin-$1/2$ lattice models \cite{San-12a,San-12b}.

Turning to interacting boson systems, thermalization was investigated by
Borgonovi et al. \cite{Borg-98} using a simple symmetrized coupled two-rotor
model. They explored different definitions of temperature and compared the occupancy
number distribution with the Bose-Einstein (BE)
distribution. They conclude that: ``For chaotic eigenstates, the distribution of
occupation numbers can be approximately described by the BE
distribution, although the system is isolated and consists of  two particles
only. In this case a strong enough interaction plays the role of a heat bath,
thus leading to thermalization". As BEGOEs
\cite{PDPK,Ch-03,Ch-04,Ma-arxiv,Ag-01} are generic models for finite isolated
interacting many-boson systems, it is important to investigate
thermalization using these ensembles.

Embedded Gaussian orthogonal ensemble of one- plus random two-body matrix ensembles for
spin-less boson systems is called BEGOE(1+2) and this ensemble was introduced and
analyzed for spectral and wave-function properties in \cite{PDPK,Ch-03,Ch-04}.
For $m$ bosons in $N$ single particle (sp) levels, in
addition to dilute limit (defined by $m \rightarrow \infty, N \rightarrow
\infty$ and $m/N \rightarrow 0$), another limiting situation, namely the dense limit (defined by $m \to \infty$, $N  \to
\infty$ and $m/N  \to \infty$) is also feasible. This limiting situation is absent for fermion
systems. Therefore the focus was on the dense limit in BEGOE investigations
\cite{PDPK,Ch-03,Ch-04,Ma-arxiv,Ag-01}.  In the strong interaction limit,
two-body part of the interaction dominates over one-body part and hence
BEGOE(1+2) reduces to BEGOE(2). Some of the generic results established for
BEGOE(1+2) are as follows: (i) eigenvalue density approaches Gaussian form
\cite{PDPK,KP-80}; (ii) for strong enough interaction, there is
average-fluctuation separation in eigenvalues \cite{PDPK,Leclair}; (iii)
similarly, the ensemble is ergodic in the dense limit with sufficiently large
$N$ \cite{Ch-03} and there will be deviations for small $N$ \cite{Ag-01}; (iv)
as the strength of the two-body interaction, $\lambda$, increases, there is
Poisson to GOE transition in  level fluctuations at $\lambda=\lambda_c$
\cite{Ch-03} and with further increase in $\lambda$, there is Breit-Wigner to
Gaussian transition in strength functions at $\lambda=\lambda_F$ \cite{Ch-04}.
The main result of the present Letter is the demonstration that finite dense
interacting boson systems generate a third chaos marker
$\lambda_t$ [as in fermionic EGOE(1+2) ensembles], a point or a region where
different definitions of entropy, temperature, specific heat and other
thermodynamic variables give the same results, i.e. where thermalization occurs.

In this Letter, we present results for thermalization in dense interacting
bosonic systems by varying the strength parameter $\lambda$ of the two-body
interaction in the BEGOE(1+2) Hamiltonian given by $H = h(1) + \lambda V(2)$. Here $h(1)$
is one-body part of the interaction, defined by single particle energies
(SPEs) $\varepsilon_k$ ($k$ = 1 to $N$) for $N$ sp levels while the
two-body interaction $V(2)$ is defined by the two-body matrix elements (TBMEs) denoted as  $V_{ijkl}=\lan(ij)|V(2)|(kl)\ran$.
 In the present study SPEs are taken as
independent gaussian random variables with mean equal to $k$ and variance equal
to $1/2$. Similarly, TBMEs are taken as independent gaussian random variables with zero
mean and variance=1 for off diagonal TBMEs and variance=2 for diagonal TBMEs.
Construction of the $m$-boson Hamiltonian, $H(m)$, and thereby the BEGOE(1+2)
ensemble in $m$-particle space with matrix dimension $d={N+m-1 \choose m}$ was
described completely in \cite{PDPK,Ch-04}. The results, presented here, have been
obtained by fully diagonalizing 100-members of a BEGOE(1+2) ensemble with 10 bosons in
5 sp levels for each value of $\lambda$. The dimensionality of the system  is $d=1001$. (we
have also carried out calculations for 10 bosons in 4 sp levels and similar results were
obtained, but they are not presented here as this is a much smaller
example, $d=286$). The ensemble average is carried out by making the spectra of each
member of the ensemble zero centered ($\epsilon$ is centroid) and scaled
to unit width ($\sigma$ is width).

The paper is organized as follows. In Section \ref{sec:sc2}, different definitions of
temperature are given and results obtained by varying the two-body interaction
strength in BEGOE(1+2) are described. Similarly, results obtained using
three different definitions for entropy, are described in Section \ref{sec:sc3}. They
allow us to define the thermalization marker $\lambda_t$. In Section \ref{sec:sc4}, duality
point is discussed, using information entropy and strength functions, in the two extreme basis
defined  by $h(1)$ and $V(2)$ operators and the $(m,N)$ dependence of the marker
$\lambda_t$ is derived. Finally, Section \ref{sec:sc5} gives conclusions.

\section{Temperature: Definitions and Results}
\label{sec:sc2}

Temperature can be defined in a number of different ways in the standard
thermodynamical treatment. These definitions of temperature are known to give
same result in the thermodynamical limit i.e. near a region where thermalization
occurs \cite{Rigol-08}.  In this section, four different definitions of temperature ($T=\beta^{-1}$), described below, have been used to compute the
temperature of finite dense interacting boson systems as a function of energy as well as a function of the two-body interaction strength.

\begin{itemize}

\item {$\beta_c$: defined using the canonical expression, between energy and temperature
which allows standard thermodynamical description for the quantum system, is given by

\be
{\lan E \ran}_{\beta_c} = \frac{\sum_i E_i\; \exp[-\beta_c\; E_i]}{
\sum_i \exp[-\beta_c\; E_i]}\;;
\ee

where $E_i$ are the eigen-energies of the Hamiltonian. With above relation,
${\lan E \ran}$ can be obtained at given $\beta_c$ using all the eigen-energies
of the system.}

\item {$\beta_{fit}$: defined using occupation numbers obtained by making use of
the standard canonical distribution is given by,

\be
{\lan n(E)_k \ran} = \frac{\sum_i n(E_i)_k\; \exp[-\beta_{fit}\; E_i]}{
\sum_i \exp[-\beta_{fit} \;E_i]}.
\ee

Here $k$ is sp level index and $i$ is eigen-energy index. Using expectation values of occupancies
calculated for all eigen-states and exact eigen-energies, $\beta_{fit}$ can be
computed by considering Eq. (2) as one parameter fitting expression and with
the constraint, $\sum_k n(E)_k=m$.}

\item  {$\beta_{BE}$: defined using BE distribution  for the
occupation numbers is given by,

\be
n(E)_k^{BE} = 1/\{\exp[\beta_{BE}(E)\;(\varepsilon_k - \mu(E))] - 1\}.
\ee

Here $\mu$ is a chemical potential. Although, this expression is derived for
many-body non-interacting particles in contact with a thermostat, it is shown
that conventional quantum statistics can appear even in isolated systems with
relatively few particles, provided a proper renormalization of energy is taken
\cite{Flam-97,Flam-96}. Comparing numerical data of expectation values of
occupancies at a particular eigenenergy and given SPEs, the unknowns $\beta_{BE}$ and
$\mu$ in the BE distribution with constraint $\sum_k
n(E)_k^{BE}=m$ can be obtained.}

\item {$\beta_T$: defined using state density, $\rho(E)$, of the total
Hamiltonian. The thermodynamic entropy is defined as
$S^{ther}(E)=\log[\rho(E)]$. The $\beta_{T}$ can be computed  using
$\beta_{T}=\frac{d \ln[\rho(E)]}{d E}$. Here not that, for BEGOE(1+2),
the form of state density is very close to Gaussian irrespective of the value of
two-body interaction strength \cite{PDPK,KP-80}.}

\end{itemize}

\begin{figure}[tbh]
 \includegraphics[width=\linewidth]{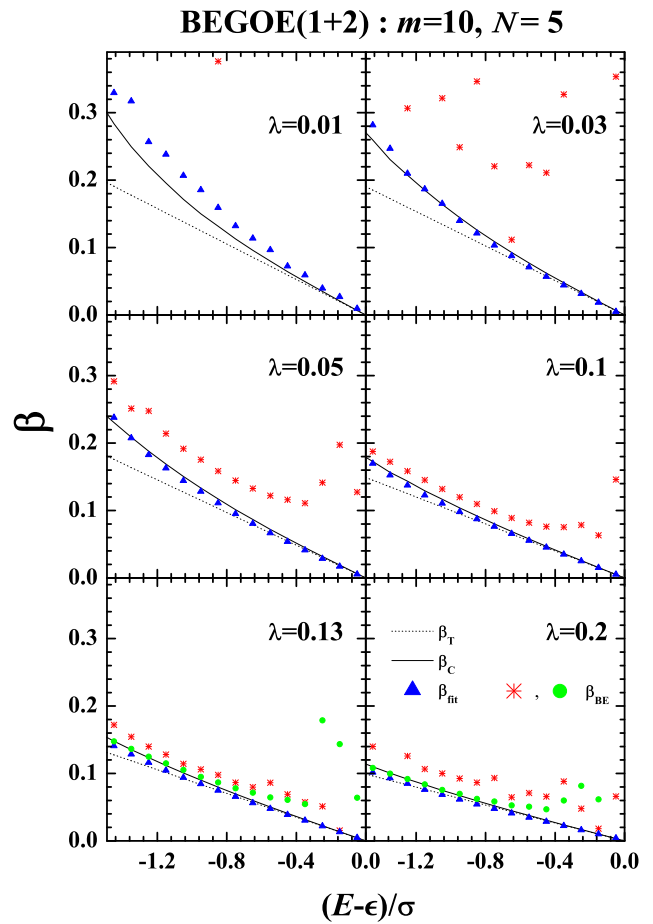}

\caption{(Color online) Ensemble averaged values of inverse temperature ($\beta$) as a
function of normalized energy, $(E-\epsilon)/\sigma$, for different values of
two body interaction strength $\lambda$, calculated using a 100 member
BEGOE(1+2) ensemble with $m=10$ and $N=5$.  Here in
calculations, different definitions of inverse of temperature are used. In the
calculations, SPEs, drawn from independent real Gaussian random variables,
are employed. Results are also shown, in the plots for $\lambda=0.13$ and $0.2$, for $\beta_{BE}$ by green filled circles, in which induced SPEs from the two-body interaction are taken into account.}

\label{fig1}
\end{figure}

Figure \ref{fig1} shows ensemble averaged values of $\beta$, computed via definitions
 described above, for a 100 member BEGOE(1+2) ensemble with $m=10$
and $N=5$ as a function of normalized energy, $\hat{E}=(E-\epsilon)/\sigma$, for
various $\lambda$ values. Here, we compare numerical values of $\beta$ from $\hat{E}=-1.5$  to
the center of the spectrum, where temperature is infinity. The edges of the
spectrum have been avoided for the following reasons:
(i) density of states is small near the edges of the spectrum; (ii) eigenstates
near edges are not fully chaotic. The $\beta$ values are obtained for all
members separately and then ensemble average is carried out taking bin-size
equal to  0.1. Since the state density for BEGOE(1+2) is Gaussian irrespective
of $\lambda$ values, $\beta_T$ as a function of energy gives straight line.
In Fig. \ref{fig1}, $\beta_T$ results are shown in the plots by dotted lines. It is
clearly seen from Fig. \ref{fig1} that for the interaction strength $\lambda < \lambda_c$ (For $(m,N)=(10,5)$,
$\lambda_c \sim 0.02$ and $\lambda_F \sim 0.05$, see ref.\cite{Ch-03}), there is significant difference between
the numerical values of $\beta$ obtained via various definitions of temperature. Going further beyond $\lambda_c$ ($\lambda_c < \lambda < \lambda_F$) where GOE fluctuations in state density sets in but the eigenstates are still not fully chaotic, the $\beta$ values
obtained via canonical expressions, defined by Eq.(1) and (2), give good agreement near the center of the spectrum. There are deviations near low temperature region. Near the region $\lambda = \lambda_F$ and beyond, the eigenstates become fully chaotic giving very good agreement between the numerical values of $\beta_c$ and $\beta_{fit}$. The inverse temperature $\beta_{BE}$ is obtained by solving microcanonical definition given by Eq. (3), with SPEs taken as independent Gaussian random variables and results are shown in the plots by red stars.  In the region $\lambda <
\lambda_F$, inverse temperature $\beta_{BE}$, found from BE distribution turns out to
be completely different from $\beta$ values obtained using other definitions. As in this region, the
structure of eigenstates is not chaotic enough, leading to strong variation in
the distribution of the occupation numbers and thus strong fluctuations in
$\beta_{BE}$. Moreover, near the center of the spectrum (i.e. as $T \to
\infty$), the value of denominator in Eq. (3) becomes very small, which
leads to large variation in $\beta_{BE}$ values from member to member. Further
increase in $\lambda > \lambda_F$, in the classically chaotic region, the
occupation number distribution becomes statistically stable with respect
to the choice of eigenstate and at one point $\lambda=\lambda_t$, temperatures defined using
canonical and microcanonical definitions give same result, i.e. $\beta_{fit}
\simeq \beta_{BE}$. This lead to same values of temperature giving the
thermodynamic marker $\lambda_t$. In the strong interaction domain ($\lambda \geq \lambda_t$), the match between $\beta_{BE}$ with other $\beta$ values is not good. This is due to neglect of induced SPEs, $\widetilde{\varepsilon_k}=\frac{m-1}{N+2} \sum_j V_{kjkj}$, from two-body interaction \cite{KP-80}. When the interaction is weak, induced SPEs part is small but in strong interaction domain their contribution is important. Adding induced SPEs part into SPEs, $\beta_{BE}$  is obtained by taking a proper renormalization of energy and results are also presented in Fig. 1 for $\lambda$ values 0.13 and 0.2 by green filled circles. Here, $\beta_{BE}$ values come quite close to other $\beta$ values. We found that the match between different values of $\beta$ is good near $\lambda=\lambda_t=0.13$.  In the next section, we
present our results for similar study using the different definitions of
entropy.

\section{Entropy: Definitions and Results}

\label{sec:sc3}

In this section, to identify the thermalization region and hence value of the
third marker $\lambda_t$ for BEGOE(1+2), we consider following three definitions
of entropy.

\begin{itemize}

\item Thermodynamic entropy, obtained using the state density $\rho(E)$, as a
function of of energy eigenvalues; $S^{ther}(E)=\log[\rho(E)]$.

\item Information entropy in the mean-field basis defined by $S^{info}(E)=-\sum_i^d{|C_i^E|^2 \log(|C_i^E|^2)}$,
 here $|C_i^E|^2$ is the
probability of basis state $i$ in the eigenstate at energy $E$.

\item Single particle entropy, obtained by calculating the occupancy of
different single-particle states, as a function of energy eigenvalues;
$S^{sp}(E)= -\sum_k{\lan n_k(E) \ran} \log({\lan n_k (E)\ran})$. Here the
summation is over all $N$ sp levels and $n_k(E)$ is the occupancy
of the $k$-th sp level at energy $E$.

\end{itemize}

\begin{figure}[tbh]
\includegraphics[width=\linewidth]{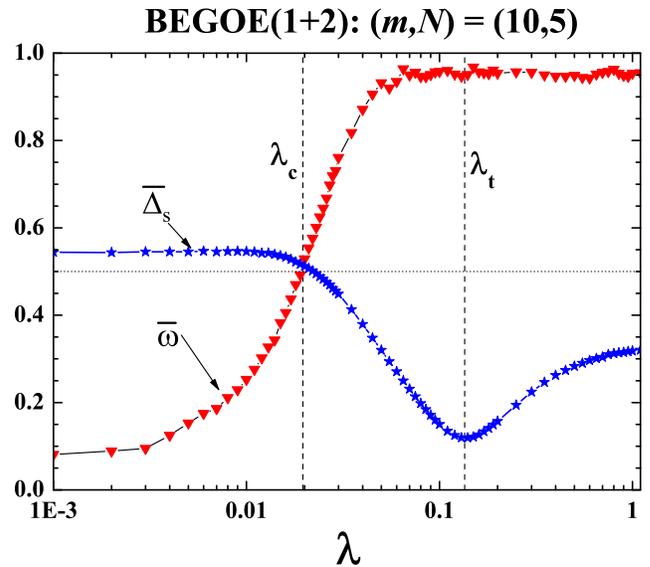}

\caption{(Color online) Ensemble averaged values $\overline{\omega}$  and
$\overline{\Delta_s}$  as a function of two body interaction strength $\lambda$,
calculated using a 100 member BEGOE(1+2) ensemble with $m=10$ and $N=5$. The vertical dash-lines represent the position of
$\lambda_c$ and $\lambda_t$. Here $\lambda_c \simeq 0.02$ and $\lambda_t \simeq 0.13$. }

\label{fig2}
\end{figure}

We use following measure, defined using above definitions of entropy
\cite{Kota-11}, to obtain $\lambda_t$:

\be
\Delta_s(\lambda)= \frac{\sqrt{ \int^{\infty}_{-\infty} [ (R^{info}_E-R^{ther}_E)^2
+ (R^{sp}_E-R^{ther}_E)^2 ] dE }} {
\int^{\infty}_{-\infty} R^{ther}_E dE },
\ee

where  $R^{\alpha}_E=\exp[S^{\alpha}(E)-S^{\alpha}_{max}]$. In the thermodynamic
region the values of the different entropies will be very close to each other,
hence the minimum of $\Delta_s$ gives the value of $\lambda_t$. In Fig. \ref{fig2},
results shown for ensemble averages $\overline{\Delta_s(\lambda)}$ (blue stars)
obtained for a $100$ member BEGOE(1+2) ensemble with $10$ bosons in $5$ sp levels
 as a function of $\lambda$. The second vertical dash-line indicates the position of $\lambda_t$ where ensemble
average $\overline{\Delta_s(\lambda)}$ is minimum. For the present example, we
obtained $\lambda_t \simeq 0.13$. This value of $\lambda_t$ is same as obtained
in Section \ref{sec:sc2}, at which different definitions of temperature give same values.

In the past it is demonstrated that for BEGOE(1+2),  as the strength $\lambda$
of the two-body interaction increases, there is Poisson to GOE transition in
level fluctuations at $\lambda=\lambda_c$ \cite{Ch-03}. In order to show that
$\lambda_c << \lambda_t$, we study the nearest neighbor spacing distribution
(NNSD) as a function of $\lambda$ to detect the position of marker $\lambda_c$.
It is well known that when the system is in integrable domain, the form of NNSD
is close to the Poisson distribution, i.e, $P(s) = \exp(-s)$, while in chaotic
domain, the form of  NNSD is the Wigner surmise, $P(s) = (\pi s/2)\; \exp (-\pi
s^2/4)$. To interpolate between these two extremes,  we use Brody distribution
\cite{Brody-81}, $P(s,\omega) = A_{\omega}(\omega + 1)s^{\omega} \exp (-
A_{\omega}\; s^{\omega+1})$. Here $\omega$ is called the Brody parameter and
$A_{\omega}$ is a normalization constant. If the spectral fluctuations are close
to the Poisson type, $\omega = 0$, or to the Wigner surmise with $\omega = 1$.
The position of the chaos marker $\lambda_c$ is fixed by the condition
$\omega(\lambda) = 1/2$. Here, the NNSD is obtained, using the unfolding
procedure described in \cite{PDPK}, with the smooth density taken as a corrected
Gaussian with corrections involving up to $6$th order moments of the density
function. In Fig. \ref{fig2}, Ensemble averaged values of $\overline{\omega(\lambda)}$
are shown by filled triangles for a $100$ member BEGOE(1+2) ensemble with
($m,N$)=($10$,$5$) as a function of two-body interaction strength $\lambda$.
The  $\lambda_c$ is shown by vertical dash line in the Fig. \ref{fig2}. Here, different criterion is used, than in \cite{Ch-03}, to obtain $\lambda_c$ although match is very good. The results
clearly show that $\lambda_c << \lambda_t$ for BEGOE(1+2) just as seen for
EGOE(1+2) fermionic ensembles \cite{Kota-11,Ma-10}.

\section{Duality and $(m,N)$ dependence of $\lambda_t$}
\label{sec:sc4}

\begin{figure}[tbh]
 \includegraphics[width=\linewidth,height=3.6in]{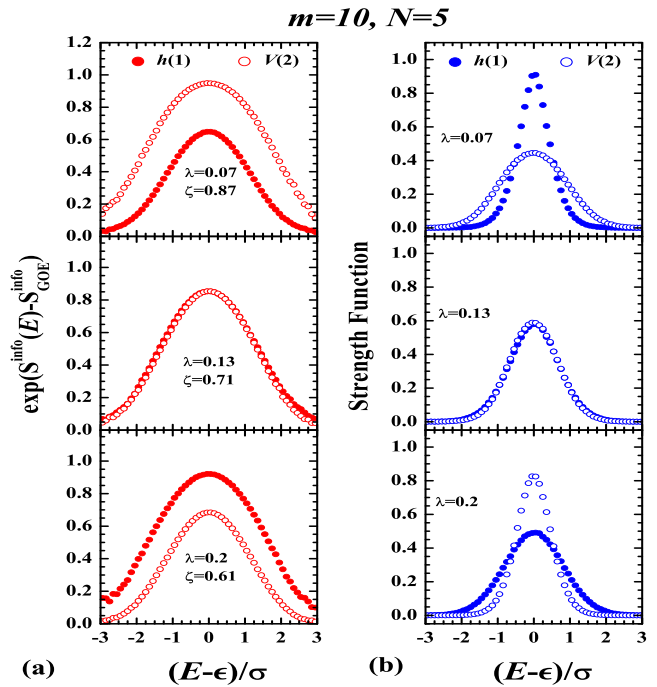}

\caption{(Color online) Ensembles averaged results for (a) the information entropy and (b) strength functions $F_{\xi_k}(E)$ in the $h(1)$ and
$V(2)$ basis for a 100 member BEGOE(1+2) ensemble with  $(m=10)$ and $(N=5)$ are shown as a
function of normalized energy, $(E-\epsilon)/\sigma$, for different $\lambda$ values. Results averaged over bin-size 0.1 are shown as circles;  filled circles correspond to $h(1)$ basis and open circles correspond to $V(2)$ basis. Ensemble averaged $\zeta$ values are
also given in the figure. Strength function plots are obtained for $\xi_k = 0$ and in the plots $\int F_{\xi_k}(E)dE=1$.}

\label{fig3}
\end{figure}

The duality region $\lambda=\lambda_d$ is the point where quantities defining
eigenstate properties like entropy, strength functions, temperature etc. give
same values irrespective of the defining basis \cite{Jacq-02}. Then one can
argue that in this region all eigenstates look alike and the duality region
defined by $\lambda=\lambda_d$ is expected to correspond to the thermodynamic
region defined by $\lambda=\lambda_t$ as in fermionic EGOE(1+2) results
\cite{An-04,Ma-10}. For the BEGOE(1+2) Hamiltonian, two choices of
basis appear naturally. One is mean-field basis defined by $h(1)$ and another is
the infinite interaction strength basis defined by $V(2)$. To examine duality,
we compare information entropy $S^{info}(E)$  and strength functions $F_{\xi_k}(E)$ (also called local density of states (LDOS))  in $h(1)$ basis and in $V(2)$ basis. The strength function corresponding to the $k$'th basis state for a particular member of ensemble is defined by $F_{\xi_k}(E)= \sum_i |C_k^i|^2 \delta(E-E_i)$, where $k$-energies $\xi_k = \langle k|H|k \rangle$ and $|k \rangle$ is the $k$'th basis state for $m$-particles in $N$ sp states. Figure \ref{fig3} shows numerical results for $S^{info}(E)$ and $F_{\xi_k}(E)$ in $h(1)$ and in $V(2)$ basis for
a 100 member BEGOE(1+2) ensemble with ($m=10$, $N=5$) for different values of
$\lambda$. Here strength functions $F_{\xi_k}(E)$ are computed following the procedure described in \cite{Ch-04} and results are shown for $\xi_k=0$ in the Fig. \ref{fig3}b. It is seen from Fig. \ref{fig3} that values of $S^{info}(E)$ and $F_{\xi_k}(E)$ in these
two basis are found very close near $\lambda=0.13$ giving value for the duality marker
$\lambda_d \simeq 0.13$ for the present example. For $\lambda < \lambda_d$, the $S^{info}(E)$ values in the $h(1)$
basis are smaller compared to those in the $V(2)$ basis and for  $\lambda >
\lambda_d$, $S^{info}$ in the $h(1)$ basis is comparatively larger. While opposite behavior is observed from the results of the strength functions. The values of entropy as well as of strength functions in these two basis coincide near $\lambda=\lambda_d$. This value is
very close to  the marker $\lambda_t$ and therefore, $\lambda_d$ region
can be interpreted as the thermodynamic region in the sense that all different
definitions of temperature and entropy coincide in this region.

In the $h(1)$ basis, $S^{info}(E)$ is determined using equation given below
\cite{Ch-03}.

\be
\exp[S^{info}(E)- S^{info}_{GOE}]= \sqrt{1-\zeta^2} \;
\exp \{\frac{\zeta^2}{2}\} \; \exp\{-\frac{\zeta^2 E^2}{2}\}\;.
\ee

Here $\zeta$ is the correlation coefficient between the full Hamiltonian $H$ and
the diagonal part of the full Hamiltonian $H$; it is given by

\be
\zeta = \sqrt{1 - \frac{\sigma^2_{off-diagonal}}{\sigma^2_{H}}} =
\sqrt{\frac{\sigma^2_{h(1)}}{\sigma^2_{h(1)}+\lambda^2 \sigma^2_{V(2)}}}\;\;.
\ee

\begin{figure}[tbh]
\includegraphics[width=\linewidth]{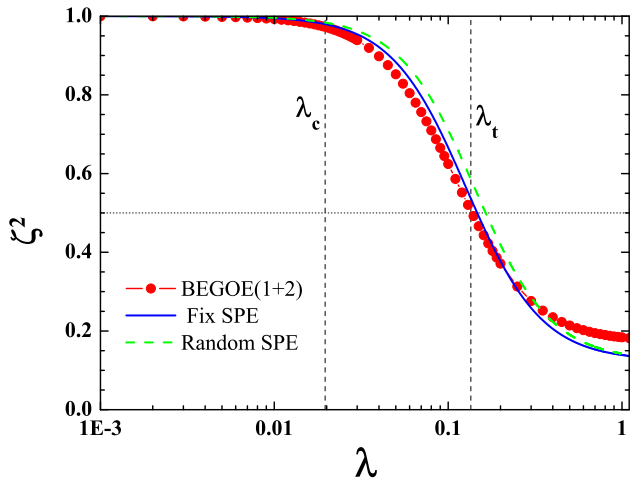}

\caption{(Color online) Ensemble averaged values of $\zeta^2$ (red filled
circles) as function of the two-body interaction strength $\lambda$,  calculated
for a 100 member BEGOE(1+2) ensemble with ($m=10, N=5$) are shown. Smooth curves are
obtained using Eq.(7) of ref.\cite{Ch-04}. Blue (continuous) curve is for fixed
SPEs and green (dash) curve is due to random SPEs used in present study.}

\label{fig4}

\end{figure}

We can determine the value of $\lambda_t$ by using the condition that
$\zeta^2=0.5$ \cite{An-04}; i.e. the spreadings produced by $h(1)$ and $V(2)$
are equal at $\lambda_t$. In Fig. \ref{fig4}, ensemble averaged values of $\zeta^2$ as a
function of $\lambda$ for a 100 member BEGOE(1+2) ensemble using ($m,N$)=($10,5$) is
presented by filled red circles. It is clear from the figure that for $\lambda
\leq \lambda_c$, $\zeta^2$ is close to $1$ and as $\lambda$ increases, $\zeta^2$
goes on decreasing smoothly. The two vertical dash-lines in Fig. \ref{fig4} indicate
the respective positions of $\lambda_c$ and $\lambda_t$ as obtained in  Section \ref{sec:sc3}.
 It can be clearly seen that $\zeta^2=0.5$ gives the thermalization point
$\lambda_t = 0.13$. For BEGOE(1+2) ensemble, analytical expression for $\zeta$
based on the method of trace propagation is derived in \cite{Ch-04}. With
$\zeta^2=0.5$ in Eq.(7) of \cite{Ch-04} and solving it for $\lambda$, ($m,N$)
dependence of marker is given by,

\be
\lambda_t = 2 \sqrt{\frac{(N+2) X}{N(N+1)(N-2)(m-1)(N+m+1)}}\;;
\label{eq.lt1}
\ee

where $X=\sum_i^N{\tilde{\varepsilon_i}}^2$. For uniform sp spectrum with
$\varepsilon_i=i$, $X=N(N+1)(N-1)/12$ and $\lambda_t =
\sqrt{\frac{(N-1)(N+2)}{3(N-2)(m-1)(N+m+1)}}$. With $(m=10,N=5)$, we have
$\lambda_t \approx 0.15$. For single particle energies which we have used in the
present study, $X=N(N^2+5)/12$ and

\be
\lambda_t =
\sqrt{\frac{(N+2)(N^2+5)}{3(N+1)(N-2)(m-1)(N+m+1)}}\;.
\label{eq.lt2}
\ee

With $(m=10,N=5)$, we have $\lambda_t \approx 0.16$. Figure 4 shows plots of $\zeta^2$ as a function of $\lambda$ obtained using Eq.(7) of
\cite{Ch-04}. The blue curve in the figure is obtained due to uniform SPEs while the green curve is obtained due to SPEs
employed in the present study. It can be seen from results that the ensemble averaged values are close to the
expected values. Small discrepancy is due to the neglect of induced single-particle energies.
 In the dense limit, Eq.(\ref{eq.lt2})
gives  $\lambda_t \sim \frac{1}{m} \sqrt{\frac{N}{3}}$. Similarly, in the dilute
limit, we have $\lambda_t \sim \frac{1}{\sqrt{3m}}$ and this result is in
agreement with EGOE(1+2) result given in \cite{An-04}. From Eq.(\ref{eq.lt2}) it is seen that for $m/N$ fixed as $m \rightarrow \infty$ and $N \rightarrow \infty$  (also into strict dense limit), $\lambda_t \rightarrow 0$. A similar behavior is expected for $\lambda_c$ and $\lambda_F$. These sudden transitions with $\lambda > 0$ are similar to the situation with Poisson to GOE or GUE \cite{KS-1999,FGM-1998} and GOE to GUE \cite{PM-1983}.

\section{Conclusions}

\label{sec:sc5}

In the present work, we have analyzed the relationship between order to chaos
transition and thermalization in finite dense interacting boson systems using
one- plus two-body embedded Gaussian orthogonal ensemble of random matrices.
Using numerical calculations, it is demonstrated that in a region $\lambda \sim
\lambda_t$ different definitions give essentially same values for temperature
and entropy, thus defining a thermalization region similar to as in fermionic
EGOE(1+2) ensembles. The value of $\lambda_t$ is much larger than the $\lambda$
value where level fluctuations change from Poisson to GOE and strength functions
from Breit-Wigner to Gaussian ($\lambda_t > \lambda_F > \lambda_c$). Further it is established that the duality region
where information entropy and strength functions will be the same in both the mean field and
interaction defined basis corresponds to a region of thermalization. We have
also obtained formula for $\lambda_t$ in terms of $(m,N)$. In addition to this, we know that
$\lambda_t >> \lambda_c$, which is further confirmed by the analytical formula for $\lambda_t$ given in Eq.(\ref{eq.lt2}) and the estimates for $\lambda_c$ as given in \cite{Ch-03}. Similar results are known for fermion where $\lambda_c \propto \frac{1}{m^2N}$ and $\lambda_t \propto \frac{1}{\sqrt{m}}$. However, for bosons formulas are not available  for $\lambda_c$ and $\lambda_F$. Therefore we cannot tell if $\lambda_F$ will be close to $\lambda_c$ or $\lambda_t$ or it will be far from both. This is an important open problem. The present work
brings completion to the study of transition (chaos) markers generated by
BEGOE(1+2) initiated in \cite{Ch-03,Ch-04}. Further investigations on
thermalization in BEGOE(1+2) will be discussed in future.

\section*{Acknowledgments}
Authors (N.D.C. and V.P.) acknowledge support from UGC(New Delhi) grant F.No:40-
425/2011(SR) and No.F.6-17/10(SA-II) respectively.





\end{document}